\documentclass[11pt, letter]{article}
\usepackage{naacl-hlt}
\usepackage{times}
\usepackage{latexsym}
\usepackage{graphicx}
\usepackage{multirow} 

\newcommand\T{\rule{0pt}{2.6ex}}

\title{How Difficult is it to Develop a Perfect Spell-checker?\\ A Cross-linguistic Analysis through Complex Network Approach}

\author{
Monojit Choudhury$^1$, Markose Thomas$^2$, Animesh Mukherjee$^1$,\\ {\bf Anupam Basu$^1$, and Niloy Ganguly$^1$}\\$^1$Department of Computer Science and Engineering, IIT Kharagpur, India\\{\normalsize \tt\{monojit,animeshm,anupam,niloy\}@cse.iitkgp.ernet.in}\\$^2$Google Inc. Bangalore, India\\{\normalsize \tt markysays@gmail.com}
}

\begin{document}
\maketitle

\begin{abstract}
The difficulties involved in spelling error detection and correction in a language have been investigated in this work through the conceptualization of SpellNet -- the weighted network of words, where edges indicate orthographic proximity between two words. We construct SpellNets for three languages - Bengali, English and Hindi. Through appropriate mathematical analysis and/or intuitive justification, we interpret the different topological metrics of SpellNet from the perspective of the issues related to spell-checking. We make many interesting observations, the most significant among them being that the probability of making a real word error in a language is propotionate to the average weighted degree of SpellNet, which is found to be highest for Hindi, followed by Bengali and English. 
\end{abstract}

%%%%%%%%%%%%%%%%%%%%%%%%%%%%%%%%%%%%%%%%%%%%%%%%%%%%%%%%%%%%%%%%%%%%%%%%%%%%%%%%%%%%%%%%%%%%%%%%%%%%%%%%%
%%%%%%%%%%%%%%%%%%%%%%%%%%%%%%%%%%%% SECTION %%%%%%%%%%%%%%%%%%%%%%%%%%%%%%%%%%%%%%%%%%%%%%%%%%%%%%%%%%%%
%%%%%%%%%%%%%%%%%%%%%%%%%%%%%%%%%%%%%%%%%%%%%%%%%%%%%%%%%%%%%%%%%%%%%%%%%%%%%%%%%%%%%%%%%%%%%%%%%%%%%%%%%
%%%%%%%%%%%%%%%%%%%%%%%%%%%%%%%%%%%%%%%%%%%%%%%%%%%%%%%%%%%%%%%%%%%%%%%%%%%%%%%%%%%%%%%%%%%%%%%%%%%%%%%%%
\section{Introduction}\label{sec:intro}  

{\em Spell-checking} is a well researched area in NLP, which deals with detection and automatic correction of spelling errors in an electronic text document. Several approaches to spell-checking have been described in the literature that use statistical, rule-based, dictionary-based or hybrid techniques (see~\cite{kukich} for a dated but substantial survey). Spelling errors are broadly classified as {\em non-word errors} (NWE) and {\em real word errors} (RWE). If the misspelt string is a valid word in the language, then it is called an RWE, else it is an NWE. For example, in English, the word ``fun" might be misspelt as ``gun" or ``vun"; while the former is an RWE, the latter is a case of NWE. It is easy to detect an NWE, but correction process is non-trivial. RWE, on the other hand are extremely difficult to detect as it requires syntactic and semantic analysis of the text, though the difficulty of correction is comparable to that of NWE (see~\cite{rwe} and references therein).

Given a lexicon of a particular language, how hard is it to develop a perfect spell-checker for that language? Since context-insensitive spell-checkers cannot detect RWE and neither they can effectively correct NWE, the difficulty in building a perfect spell-checker, therefore, is reflected by quantities such as the probability of a misspelling being RWE, probability of more than one word being orthographically closer to an NWE, and so on. In this work, we make an attempt to understand and formalize some of these issues related to the challenges of spell-checking through a complex network approach (see~\cite{barabasi,newman} for a review of the field).  This in turn allows us to provide language-specific quantitative bounds on the performance level of spell-checkers.

In order to formally represent the orthographic structure (spelling conventions) of a language, we conceptualize
the lexicon as a weighted network, where the nodes represent the words and the weights of the edges indicate the orthoraphic similarity between the pair of nodes (read words) they connect. We shall call this network the {\bf S}pelling {\bf N}etwork or {\bf SpellNet} for short. We build the SpellNets for three languages --  Bengali, English and Hindi, and carry out standard topological analysis of the networks following complex network theory. Through appropriate mathematical analysis and/or intuitive justification, we interpret the different topological metrics of SpellNet from the perspective of difficulties related to spell-checking. Finally, we make several cross-linguistic observations, both invariances and variances, revealing quite a few interesting facts. For example, we see that among the three languages studied, the probability of RWE is highest in Hindi followed by Bengali and English. A similar observation has been previously reported in~\cite{simple} for RWEs in Bengali and English.

Apart from providing insight into spell-checking, the complex structure of SpellNet also reveals the self-organization and evolutionary dynamics underlying the orthographic properties of natural languages. In recent times, complex networks have been successfully employed to model and explain the structure and organization of several natural and social phenomena, such as the foodweb, protien interaction, formation of language inventories~\cite{monojit}, syntactic structure of languages~\cite{cancho}, WWW, social collaboration, scientific citations and many more (see~\cite{barabasi,newman} and references therein). This work is inspired by the aforementioned models, and more specifically a couple of similar works on {\em phonological neighbors'} network of words~\cite{kapatsinski,vitevitch}, which try to explain the human perceptual and cognitive processes in terms of the organization of the mental lexicon. 

The rest of the paper is organized as follows. Section~\ref{sec:spellnet} defines the structure and construction procedure of SpellNet. Section~\ref{sec:dd} and~\ref{sec:cc} describes the degree and clustering related properties of SpellNet and their significance in the context of spell-checking, respectively. Section~\ref{sec:con} summarizes the findings and discusses possible directions for future work. The derivation of the probability of RWE in a language is presented in Appendix A.

%%%%%%%%%%%%%%%%%%%%%%%%%%%%%%%%%%%%%%%%%%%%%%%%%%%%%%%%%%%%%%%%%%%%%%%%%%%%%%%%%%%%%%%%%%%%%%%%%%%%%%%%%
%%%%%%%%%%%%%%%%%%%%%%%%%%%%%%%%%%%% SECTION %%%%%%%%%%%%%%%%%%%%%%%%%%%%%%%%%%%%%%%%%%%%%%%%%%%%%%%%%%%%
%%%%%%%%%%%%%%%%%%%%%%%%%%%%%%%%%%%%%%%%%%%%%%%%%%%%%%%%%%%%%%%%%%%%%%%%%%%%%%%%%%%%%%%%%%%%%%%%%%%%%%%%%
%%%%%%%%%%%%%%%%%%%%%%%%%%%%%%%%%%%%%%%%%%%%%%%%%%%%%%%%%%%%%%%%%%%%%%%%%%%%%%%%%%%%%%%%%%%%%%%%%%%%%%%%%
 
\section{SpellNet: Definition and Construction}\label{sec:spellnet}

\begin{figure}
\centering
\includegraphics[width=3in]{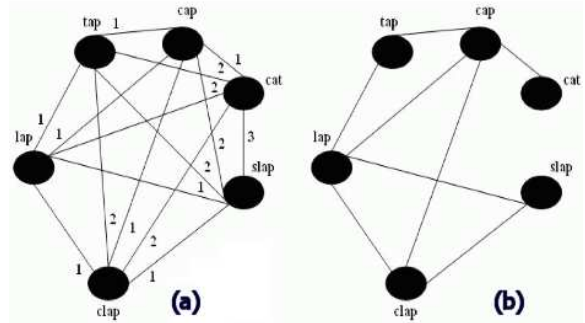}
\caption{The structure of SpellNet: (a) the weighted SpellNet for 6 English words, (b) Thresholded counterpart of (a), for $\theta=1$}
\label{fig:spellnet}
\end{figure}

In order to study and formalize the orthographic characteristics of a language, we model the lexicon $\Lambda$ of the language as an undirected and fully connected weighted graph $G (V,E)$. Each word $w \in \Lambda$ is represented by a vertex $v_w \in V$, and for every pair of vertices $v_w$ and $v_{w'}$ in $V$, there is an edge $(v_w,v_{w'}) \in E$. The {\em weight} of the edge $(v_w,v_{w'})$, is equal to $ed(w,w')$ -- the orthographic edit distance between $w$ and $w'$ (considering substitution, deletion and insertion to have a cost of 1). Each node $v_w \in V$ is also assigned a node weight $W_V(v_w)$ equal to the unigram occurrence frequency of the word $w$. We shall refer to the graph $G (V,E)$ as the SpellNet. Figure~\ref{fig:spellnet}(a) shows a hypothetical SpellNet for 6 common English words.

We define unweighted versions of the graph $G (V,E)$ through the concept of thresholding as described below. For a threshold $\theta$, the graph $G_\theta (V,E_\theta)$ is an unweighted sub-graph of $G (V,E)$, where an edge $(v_w,v_{w'}) \in E$ is assigned a weight 1 in $E_\theta$ if and only if the weight of the edge is less than or equal to $\theta$, else it is assigned a weight 0. In other words, $E_\theta$ consists of only those edges in $E$ whose edge weight is less than or equal to $\theta$. Note that all the edges in $E_\theta$ are unweighted. Figure~\ref{fig:spellnet}(b) shows the thresholded SpellNet shown in~\ref{fig:spellnet}(a) for $\theta=1$.

%%%%%%%%%%%%%%%%%%%%%%%%%%%%%%%%%%%%%%%%%%%%%%%%%%%%%%%%%%%%%%%%%%%%%%%%%%%%%%%%%%%%%%%%%%%%%%%%%%%%%%%%%
%%%%%%%%%%%%%%%%%%%%%%%%%%%%%%%%%%%%%%%%%%%%%%%%%%%%%%%%%%%%%%%%%%%%%%%%%%%%%%%%%%%%%%%%%%%%%%%%%%%%%%%%%

\subsection{Construction of SpellNets}

We construct the SpellNets for three languages -- Bengali, English and Hindi. While the two Indian languages --  Bengali and Hindi -- use Brahmi derived scripts -- Bengali and Devanagari respectively, English uses the Roman script. Moreover, the orthography of the two Indian languages are highly phonemic in nature, in contrast to the morpheme-based orthography of English. Another point of disparity lies in the fact that while the English alphabet consists of 26 characters, the alphabet size of both Hindi and Bengali is around 50.

The lexica for the three languages have been taken from public sources. For English it has been obtained from the website
{\em www.audiencedialogue.org/susteng.html}; for Hindi and Bengali, the word lists as well as the unigram frequencies have been estimated from the  monolingual corpora published by Central Institute of Indian Languages. We chose to work with the most frequent 10000 words, as the medium size of the two Indian language corpora (around 3M words each) does not provide sufficient data for estimation of the unigram frequencies of a large number of words (say 50000). Therefore, all the results described in this work pertain to the SpellNets corresponding to the most frequent 10000 words. However, we believe that the trends observed do not reverse as we increase the size of the networks.
 
In this paper, we focus on the networks at three different thresholds, that is for $\theta =1,3,5$, and study the properties of $G_\theta$ for the three languages. We do not go for higher thresholds as the networks become completely connected at $\theta = 5$. Table~\ref{tab:val} reports the values of different topological metrics of the SpellNets for the three languages at three thresholds. In the following two sections, we describe in detail some of the topological properties of SpellNet, their implications to spell-checking, and observations in the three languages.

\begin{table*}[!t]
	\centering
\resizebox*{!}{2in}{
	\begin{tabular}{|c|ccc|ccc|ccc|}
		\hline
		 &\multicolumn{3}{|c|}{\multirow{2}{*}{English}} & 
		\multicolumn{3}{|c|}{\multirow{2}{*}{Hindi}} & 
		\multicolumn{3}{|c|}{\multirow{2}{*}{Bengali}} \\ 
		&\multicolumn{3}{|c|}{} &
		\multicolumn{3}{|c|}{} &
		\multicolumn{3}{|c|}{} \\
		\cline{2-10}
		\T & $\theta=1$ & $\theta=3$ & $\theta=5$ & $\theta=1$ & $\theta=3$ & $\theta=5$ & $\theta=1$ & $\theta=3$ & $\theta=5$ \\ 
		\hline
		\hline
		%& & & & & & & & & & & &\\
		\T $M$ 
		& 8.97k& 0.70M & 8.46M& 17.6k& 1.73M& 17.1M& 11.9k & 1.11M& 13.2M\\
		%& & & & & & & & & \\
		\T $\langle k\rangle$ 
		& 2.79& 140.25& 1692.65& 4.52& 347.93 & 3440.06& 3.38& 223.72 &2640.11 \\
		%& & & & & & & & & \\
		$\langle k_{wt}\rangle$ 
		& 6.81& 408.03& 1812.56& 13.45& 751.24& 4629.36& 7.73 & 447.16& 3645.37 \\
		& & & & & & & & & \\
		$r_{dd}$ 
		& 0.696 & 0.480& 0.289& 0.696& 0.364& 0.129& 0.702 & 0.389& 0.155 \\
		$\langle CC\rangle$ 
		& 0.101 & 0.340& 0.563& 0.172& 0.400& 0.697& 0.131& 0.381& 0.645  \\
		%& & & & & & & & & \\
		$\langle CC_{wt}\rangle$ 
		& 0.221& 0.412& 0.680& 0.341& 0.436& 0.760& 0.229& 0.418& 0.681 \\
		& & & & & & & & & \\
		$\langle l\rangle$
		&7.07 & 3.50& N.E & 7.47&2.74 & N.E & 8.19&2.95 & N.E \\
		%& & & & & & & & & \\
		$D$
		&24 & 14& N.E & 26& 12& N.E & 29& 12 & N.E \\
		\hline
	\end{tabular}
	}
	\caption{Various topological metrics and their associated values for the SpellNets of the three languages at thresholds 	1, 3 and 5. Metrics: $M$ -- number of edges; $\langle k\rangle$ -- average degree; $\langle k_{wt}\rangle$ -- average weighted degree; $\langle CC\rangle$ -- average clustering coefficient; $\langle CC_{wt}\rangle$ - average weighted clustering coefficient; $r_{dd}$ -- Pearson correlation coefficient between degrees of neighbors; $\langle l\rangle$ -- average shortest path; $D$ -- diameter. N.E -- Not Estimated. See the text for further details on definition, computation and significance of the metrics.}
\label{tab:val}
\end{table*}

%%%%%%%%%%%%%%%%%%%%%%%%%%%%%%%%%%%%%%%%%%%%%%%%%%%%%%%%%%%%%%%%%%%%%%%%%%%%%%%%%%%%%%%%%%%%%%%%%%%%%%%%%
%%%%%%%%%%%%%%%%%%%%%%%%%%%%%%%%%%%% SECTION %%%%%%%%%%%%%%%%%%%%%%%%%%%%%%%%%%%%%%%%%%%%%%%%%%%%%%%%%%%%
%%%%%%%%%%%%%%%%%%%%%%%%%%%%%%%%%%%%%%%%%%%%%%%%%%%%%%%%%%%%%%%%%%%%%%%%%%%%%%%%%%%%%%%%%%%%%%%%%%%%%%%%%
%%%%%%%%%%%%%%%%%%%%%%%%%%%%%%%%%%%%%%%%%%%%%%%%%%%%%%%%%%%%%%%%%%%%%%%%%%%%%%%%%%%%%%%%%%%%%%%%%%%%%%%%%

\section{Degree Distribution} \label{sec:dd}
The {\em degree} of a vertex in a network is the number of edges incident on that vertex. 
Let $P_k$ be the probability that a randomly chosen vertex has degree $k$ or more than $k$. A plot of $P_k$ for any given network can be formed by making a histogram of the degrees of the vertices, and this plot is known as the {\em cumulative degree distribution} of the network~\cite{newman}. The (cumulative) degree distribution of a network provides important insights into the topological properties of the network.

\begin{figure*}
\centering
\includegraphics[width=2in]{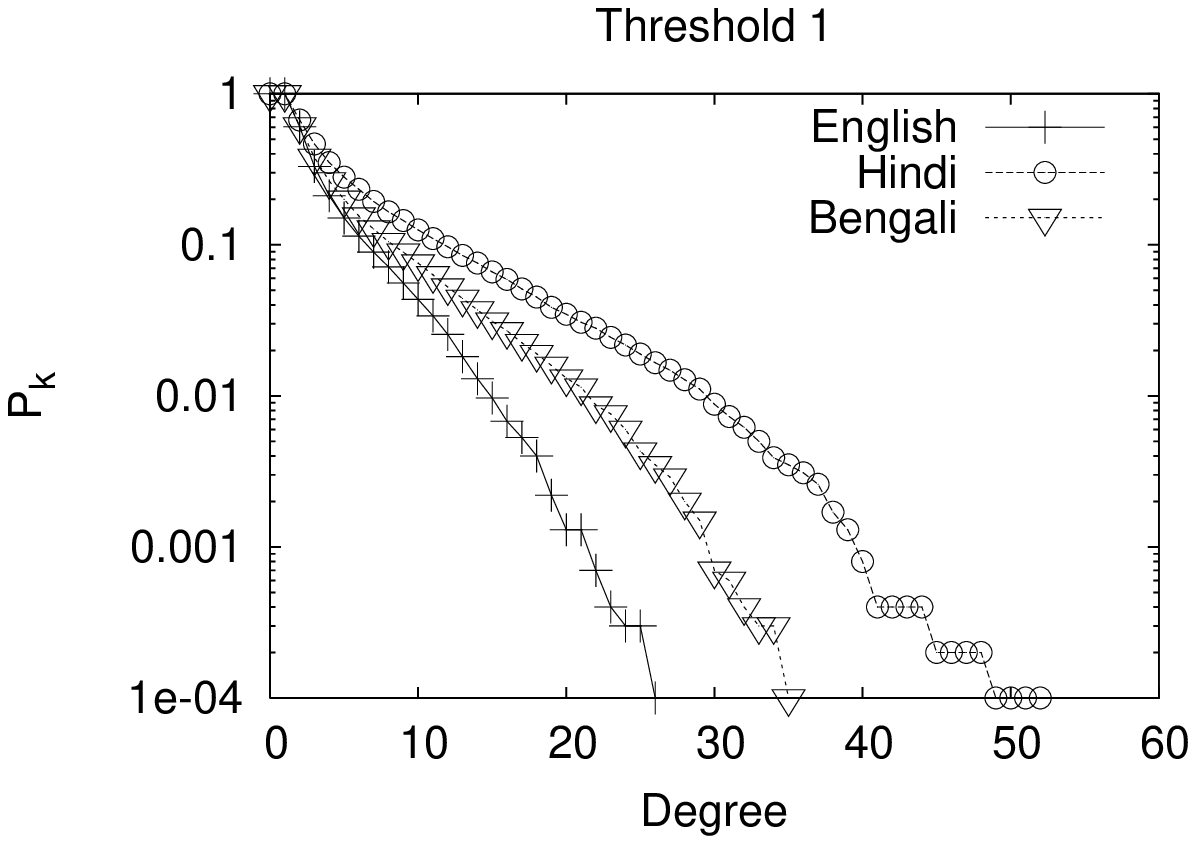}
\includegraphics[width=2in]{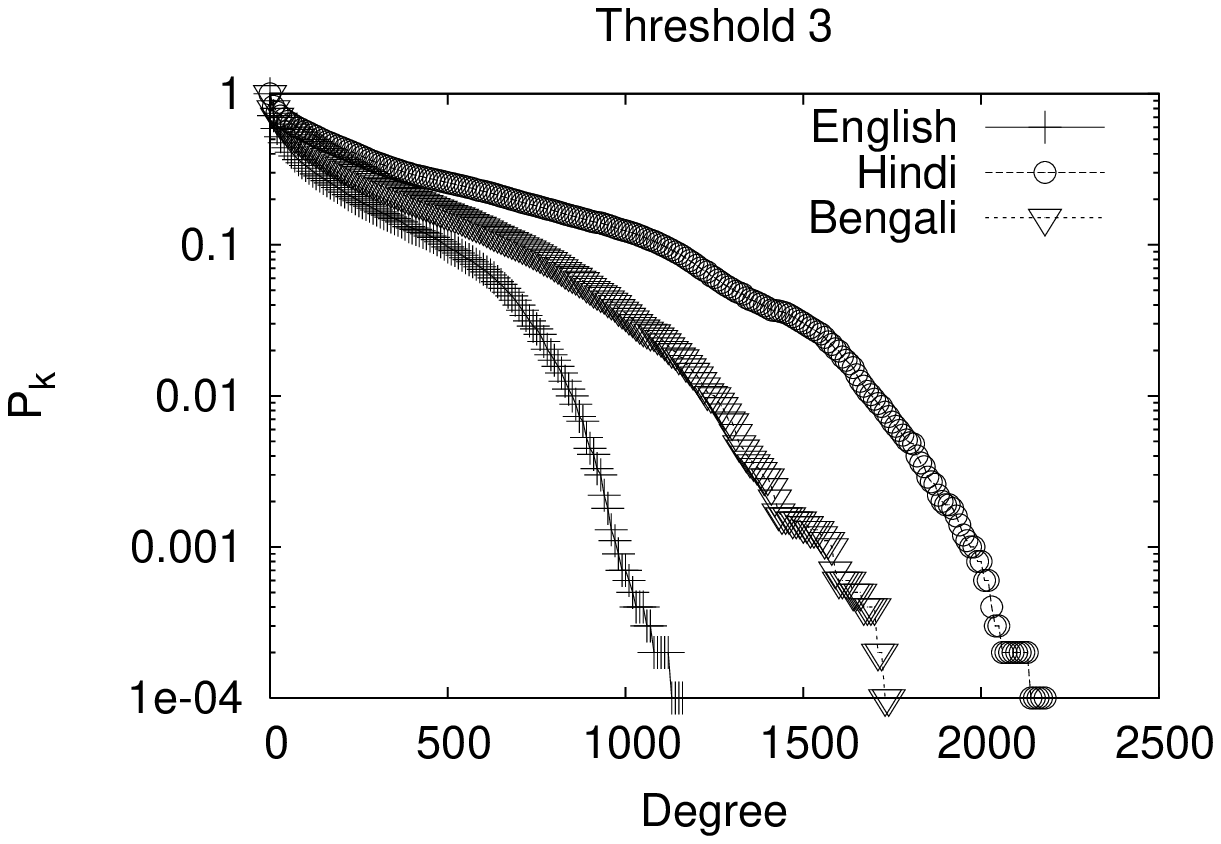}
\includegraphics[width=2in]{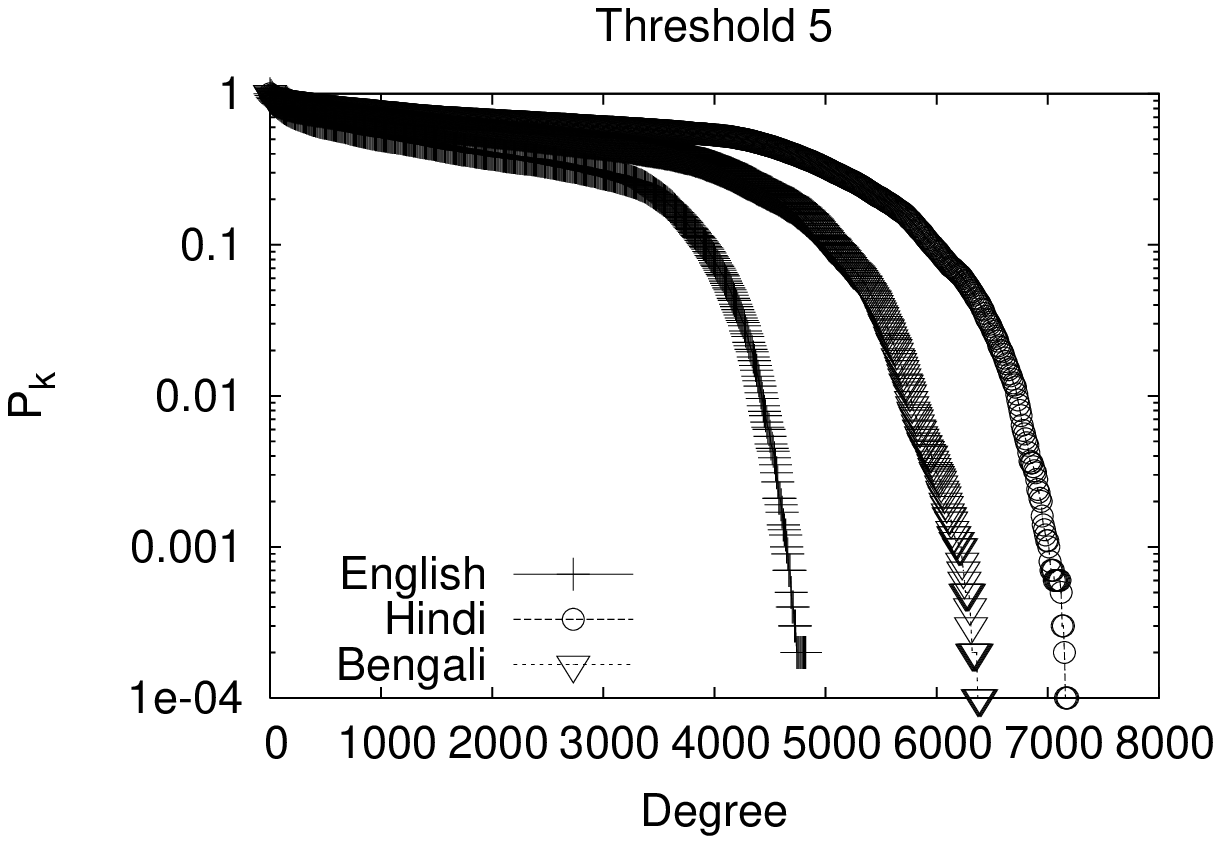}
\caption{Cumulative degree distribution of SpellNets at different thresholds presented in semi-logarithmic scale.}
\label{fig:cdd}
\end{figure*}

Figure~\ref{fig:cdd} shows the plots for the cumulative degree distribution for $\theta=1,3,5$, plotted on a 
log-linear scale. The linear nature of the curves in the semi-logarithmic scale indicates that the distribution is exponential in nature. The exponential behaviour is clearly visible for $\theta=1$, however at higher thresholds, there are very few nodes in the network with low degrees, and therefore only the tail of the curve shows a pure exponential behavior. We also observe that the steepness (i.e. slope) of the $\log(P_k)$ with respect to $k$ increases with $\theta$. It is interesting to note that although most of the naturally and socially occurring networks exhibit a power-law degree distribution (see~\cite{barabasi,newman,cancho,monojit} and references therein), SpellNets feature exponential degree distribution. Nevertheless, similar results have also been reported for the phonological neighbors' network~\cite{kapatsinski}.  

%%%%%%%%%%%%%%%%%%%%%%%%%%%%%%%%%%%%%%%%%%%%%%%%%%%%%%%%%%%%%%%%%%%%%%%%%%%%%%%%%%%%%%%%%%%%%%%%%%%%%%%%%
%%%%%%%%%%%%%%%%%%%%%%%%%%%%%%%%%%%%%%%%%%%%%%%%%%%%%%%%%%%%%%%%%%%%%%%%%%%%%%%%%%%%%%%%%%%%%%%%%%%%%%%%%

\subsection{Average Degree}

Let the degree of the node $v$ be denoted by $k(v)$. We define the quantities -- the average degree $\langle k \rangle$ and the weighted average degree $\langle k_{wt} \rangle $ for a given network as follows (we drop the subscript $w$ for clarity of notation).
\begin{equation}\label{mean_deg}
\langle k \rangle = \frac{1}{N}\sum_{v \in V}k(v)
\end{equation}  
\begin{equation}\label{wt_deg}
\langle k_{wt} \rangle = \frac{\sum_{v \in V}k(v)W_V(v)}{\sum_{v \in V}W_V(v)}
\end{equation}  
where $N$ is the number of nodes in the network. 

%%%%%%%%%%%%%%%%%%%%%%%%%%%%%%%%%%%%%%%%%%%%%%%%%%%%%%%%%%%%%%%%%%%%%%%%%%%%%%%%%%%%%%%%%%%%%%%%%%%%%%%%%

{\bf Implication:} The average weighted degree of SpellNet can be interpreted as the probability of RWE in a language. This correlation can be derived as follows. Given a lexicon $\Lambda$ of a language, it can be shown that the probability of RWE in a language, denoted by $p_{rwe}(\Lambda)$ is given by the following equation (see Appendix A for the derivation)
\begin{equation}\label{rwe3}
p_{rwe}(\Lambda) = \sum_{w \in \Lambda}\sum_{\begin{array}{c}^{w' \in \Lambda}\\ ^{w \ne w'}\end{array}}\rho^{ed(w,w')}p(w)
\end{equation}

Let $neighbor(w,d)$ be the number of words in $\Lambda$ whose edit distance from $w$ is $d$. Eqn~\ref{rwe3} can be rewritten in terms of $neighbor(w,d)$ as follows.
\begin{equation}
p_{rwe}(\Lambda) = \sum_{w \in \Lambda}\sum_{d = 1}^{\infty}\rho^d\; neighbor(w,d)p(w)
\end{equation}
Practically, we can always assume that $d$ is bounded by a small positive integer. In other words, the number of errors simultaneously made on a word is always small (usually assumed to be 1 or a slowly growing function of the word length~\cite{kukich}). Let us denote this bound by $\theta$. Therefore, 
\begin{equation}\label{rwe4}
p_{rwe}(\Lambda) \approx \sum_{w \in \Lambda}\sum_{d = 1}^{\theta}\rho^d\; neighbor(w,d)p(w)
\end{equation}

Since $\rho <1$, we can substitute $\rho^d$ by $\rho$ to get an upper bound on $p_{rwe}(\Lambda)$, which gives
\begin{equation}\label{rwe5}
p_{rwe}(\Lambda) < \rho\sum_{w \in \Lambda}\sum_{d = 1}^{\theta}neighbor(w,d)p(w) 
\end{equation}

The term $\sum_{d = 1}^{\theta}neighbor(w,d)$ computes the number of words in the lexicon, whose edit distance from $w$ is atmost $\theta$. This is nothing but $k(v_w)$, i.e. the degree of the node $v_w$, in $G_\theta$. Moreover, the term $p(w)$ is proportionate to the node weight $W_V(v_w)$. Thus, rewriting Eqn~\ref{rwe5} in terms of the network parameters for $G_\theta$, we get (subscript $w$ is dropped for clarity)
\begin{equation}\label{rwe6}
p_{rwe}(\Lambda) < \rho\frac{\sum_{v \in V}k(v)W_V(v)}{\sum_{v \in V}W_V(v)}
\end{equation}
Comparing Eqn~\ref{wt_deg} with the above equation, we can directly obtain the relation 
\begin{equation}\label{imp1}
p_{rwe}(\Lambda) < C_1\langle k_{wt}\rangle
\end{equation}
where $C_1$ is some constant of proportionality. Note that for $\theta =1$, $p_{rwe}(\Lambda) \propto \langle k_{wt}\rangle$. If we ignore the distribution of the words, that is if we assume $p(w) = 1/N$, then $p_{rwe}(\Lambda) \propto \langle k \rangle$.  

Thus, the quantity $\langle k_{wt}\rangle$ provides a good estimate of the probability of RWE in a language. 
%%%%%%%%%%%%%%%%%%%%%%%%%%%%%%%%%%%%%%%%%%%%%%%%%%%%%%%%%%%%%%%%%%%%%%%%%%%%%%%%%%%%%%%%%%%%%%%%%%%%%%%%%%%%%%%%%%%%%%%

{\bf Observations and Inference:} At $\theta=1$, the average weighted degrees for Hindi, Bengali and English are 13.81, 7.73 and 6.61 respectively. Thus, the probability of RWE in Hindi is significantly higher than that of Bengali, which in turn is higher than that of English~\cite{simple}. Similar trends are observed at all the thresholds for both $\langle k_{wt} \rangle$ and $\langle k \rangle$. This is also evident from Figures~\ref{fig:cdd}, which show the distribution of Hindi to lie above that of Bengali, which lies above English (for all thresholds). 

\begin{figure}
	\centering
	\includegraphics[scale=0.5]{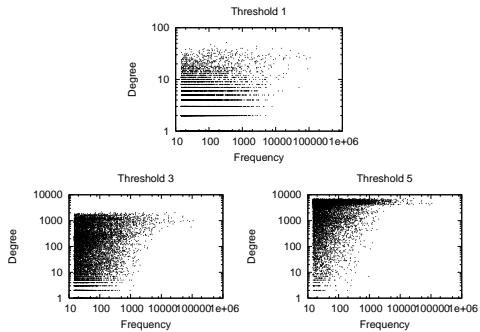}
	\caption{Scatter-plots for degree versus unigram frequency at different $\theta$ for Hindi \label{fig:degfreq}}
\end{figure}

The average degree $\langle k \rangle$ is substantially  smaller (0.5 to 0.33 times) than the average weighted degree $\langle k_{wt} \rangle$ for all the 9 SpellNets. This suggests that the higher degree nodes in SpellNet have higher node weight (i.e. occurrence frequency). Indeed, as shown in Figure~\ref{fig:degfreq} for Hindi, the high unigram frequency of a node implies higher degree, though the reverse is not true. The scatter-plots for the other languages are similar in nature. 

%%%%%%%%%%%%%%%%%%%%%%%%%%%%%%%%%%%%%%%%%%%%%%%%%%%%%%%%%%%%%%%%%%%%%%%%%%%%%%%%%%%%%%%%%%%%%%%%%%%%%%%%%
%%%%%%%%%%%%%%%%%%%%%%%%%%%%%%%%%%%%%%%%%%%%%%%%%%%%%%%%%%%%%%%%%%%%%%%%%%%%%%%%%%%%%%%%%%%%%%%%%%%%%%%%%

\subsection{Correlation between Degrees of Neighbors}
The relation between the degrees of adjacent words is described by the {\em degree assortativity coefficient}. One way to define the assortativity of a network is through the Pearson correlation coefficient between the degrees of the two vertices connected by an edge. Each edge $(u,v)$ in the network adds a data item corresponding to the degrees of $u$ and $v$ to two data sets $x$ and $y$ respectively. The Pearson correlation coefficient for the data sets $x$ and $y$ 
of $n$ items each is then defined as
\begin{displaymath}
	r=\frac{n\sum xy - \sum x \sum y}{\sqrt{[n \sum x^2 - (\sum x)^2][n \sum y^2 - (\sum y)^2]}}
\end{displaymath}

{\bf Observation:} $r$ is positive for the networks in which words tend to associate with other words of similar degree (i.e. high degree with high degree and vice versa), and it is negative for networks in which words associate with words having degrees in the opposite spectrum. Refering to table~\ref{tab:val}, we see that the correlation coefficient $r_{dd}$ is roughly the same and equal to around 0.7 for all languages at $\theta=1$. As $\theta$ increases, the correlation decreases as expected, due to the addition of edges between dissimilar words. 

%%%%%%%%%%%%%%%%%%%%%%%%%%%%%%%%%%%%%%%%%%%%%%%%%%% EDIT from HERE %%%%%%%%%%%%%%%%%%%%%%%%%%%%%%%%%%%%%%%%%%%%%%%%%%%%%%
{\bf Implication:} The high positive correlation coefficients suggest that SpellNets feature assortative mixing of nodes in terms of degrees. If there is an RWE corresponding to a high degree node $v_w$, then due to the assortative mixing of nodes, the misspelling $w'$ obtained from $w$, is also expected to have a high degree. Since $w'$ has a high degree, even after detection of the fact that $w'$ is a misspelling, choosing the right suggestion (i.e. $w$) is extremely difficult unless the linguistic context of the word is taken into account. Thus, more often than not it is difficult to correct an RWE, even after successful detection.

%%%%%%%%%%%%%%%%%%%%%%%%%%%%%%%%%%%%%%%%%%%%%%%%%%%%%%%%%%%%%%%%%%%%%%%%%%%%%%%%%%%%%%%%%%%%%%%%%%%%%%%%%
%%%%%%%%%%%%%%%%%%%%%%%%%%%%%%%%%%%% SECTION %%%%%%%%%%%%%%%%%%%%%%%%%%%%%%%%%%%%%%%%%%%%%%%%%%%%%%%%%%%%
%%%%%%%%%%%%%%%%%%%%%%%%%%%%%%%%%%%%%%%%%%%%%%%%%%%%%%%%%%%%%%%%%%%%%%%%%%%%%%%%%%%%%%%%%%%%%%%%%%%%%%%%%
%%%%%%%%%%%%%%%%%%%%%%%%%%%%%%%%%%%%%%%%%%%%%%%%%%%%%%%%%%%%%%%%%%%%%%%%%%%%%%%%%%%%%%%%%%%%%%%%%%%%%%%%%

\section{Clustering and Small World Properties}\label{sec:cc}
In the previous section, we looked at some of the degree based features of SpellNets. These features provide us insights regarding the probability of RWE in a language and the level of difficulty in correcting the same. In this section, we discuss some of the other characteristics of SpellNets that are useful in predicting the difficulty of non-word error correction. 

%%%%%%%%%%%%%%%%%%%%%%%%%%%%%%%%%%%%%%%%%%%%%%%%%%%%%%%%%%%%%%%%%%%%%%%%%%%%%%%%%%%%%%%%%%%%%%%%%%%%%%%%%
%%%%%%%%%%%%%%%%%%%%%%%%%%%%%%%%%%%%%%%%%%%%%%%%%%%%%%%%%%%%%%%%%%%%%%%%%%%%%%%%%%%%%%%%%%%%%%%%%%%%%%%%%

\subsection{Clustering Coefficient}

Recall that in the presence of a complete list of valid words in a language, detection of NWE is a trivial task. However, correction of NWE is far from trivial. Spell-checkers usually generate a suggestion list of possible candidate words that are within a small edit distance of the misspelling. Thus, correction becomes hard as the number of words within a given edit distance from the misspelling increases. Suppose that a word $w \in \Lambda$ is transformed into $w'$ due to some typing error, such that $w' \notin \Lambda$. Also assume that $ed(w,w') \le \theta$. We want to estimate the number of words in $\Lambda$ that are within an edit distance $\theta$ of $w'$. In other words we are interested in finding out the degree of the node $v_{w'}$ in $G_\theta$, but since there is no such node in SpellNet, we cannot compute this quantity directly. Nevertheless, we can provide an approximate estimate of the same as follows.

Let us conceive of a hypothetical node $v_{w'}$. By definition of SpellNet, there should be an edge connecting $v_{w'}$ and $v_w$ in $G_\theta$. A crude estimate of $k(v_{w'})$ can be $\langle k_{wt}\rangle$ of $G_\theta$. Due to the assortative nature of the network, we expect to see a high correlation between the values of $k(v_w)$ and $k(v_{w'})$, and therefore, a slightly better estimate of $k(v_{w'})$ could be $k(v_w)$. However, as $v_{w'}$ is not a part of the network, it's behavior in SpellNet may not resemble that of a real node, and such estimates can be grossly erroneous. 

One way to circumvent this problem is to look at the local neighborhood of the node $v_w$. Let us ask the question -- what is the probability that two randomly chosen neighbors of $v_w$ in $G_\theta$ are connected to each other? If this probability is high, then we can expect the local neighborhood of $v_w$ to be dense in the sense that almost all the neighbors of $v_w$ are connected to each other forming a {\em clique-like} local structure. Since $v_{w'}$ is a neighbor of $v_w$, it is a part of this dense cluster, and therefore, its degree $k(v_{w'})$ is of the order of $k(v_w)$. On the other hand, if this probability is low, then even if $k(v_w)$ is high, the space around $v_w$ is sparse, and the local neighborhood is {\em star-like}. In such a situation, we expect $k(v_{w'})$ to be low.

The topological property that measures the probability of the neighbors of a node being connected is called the {\em clustering coefficient} (CC). One of the ways to define the clustering coefficient $C(v)$ for a vertex $v$ in a network is
\begin{displaymath}
	C(v)=\frac{\textrm{number of triangles connected to vertex } v}
	{\textrm{number of triplets centered on } v}
\end{displaymath}
For vertices with degree 0 or 1, we put $C(v)=0$. Then the clustering coefficient for the whole 
network $\langle CC\rangle$ is the mean CC of the nodes in the network. A corresponding weighted version of the CC $\langle CC_{wt}\rangle$ can be defined by taking the node weights into account. 

{\bf Implication:} The higher the value of $k(v_w)C(v_w)$ for a node, the higher is the probability that an NWE made while typing $w$ is hard to correct due to the presence of a large number of orthographic neighbors of the misspelling. Therefore, in a way $\langle CC_{wt}\rangle$ reflects the level of difficulty in correcting NWE for the language in general.

{\bf Observation and Inference:} At threshold 1, the values of $\langle CC\rangle$ as well as $\langle CC_{wt}\rangle$ is higher for Hindi (0.172 and 0.341 respectively) and Bengali (0.131 and 0.229 respectively) than that of English (0.101 and 0.221 respectively), though for higher thresholds, the difference between the CC for the languages reduces. This observation further strengthens our claim that the level of difficulty in spelling error detection and correction are language dependent, and for the three languages studied, it is hardest for Hindi, followed by Bengali and English.

\subsection{Small World Property}
As an aside, it is interesting to see whether the SpellNets exhibit the so called {\em small world effect} that is prevalent in many social and natural systems~(see \cite{barabasi,newman} for definition and examles). A network is said to be a {\em small world} if it has a high clustering coefficient and if the average shortest path between any two nodes of the network is small. 

{\bf Observation:} We observe that SpellNets indeed feature a high CC that grows with the threshold. The average shortest path, denoted by $\langle l \rangle$ in Table~\ref{tab:val}, for $\theta = 1$ is around 7 for all the languages, and reduces to around 3 for $\theta = 3$; at $\theta=5$ the networks are near-cliques. Thus, SpellNet is a small world network.

{\bf Implication:}  By the application of triangle inequality of edit distance, it can be easily shown that $\langle l \rangle\times\theta$ provides an upper bound on the average edit distance between all pairs of the words in the lexicon. Thus, a small world network, which implies a small $\langle l \rangle$, in turn implies that as we increase the error bound (i.e. $\theta$), the number of edges increases sharply in the network and soon the network becomes fully connected. Therefore, it becomes increasingly more difficult to correct or detect the errors, as any word can be a possible suggestion for any misspelling. In fact this is independently observed through the exponential rise in $M$ -- the number of edges, and fall in $\langle l \rangle$ as we increase $\theta$.

{\bf Inference:} It is impossible to correct very noisy texts, where the nature of the noise is random and words are distorted by a large edit distance (say 3 or more).

%%There is another interesting interpretation of $\langle l \rangle$, and $D$ (the diameter or the longest shortest ath of %%the network). Suppose we want to convert a given word $w$ to another word $w'$ using a set of edit operations (insertion, %%deletion, substitution), such that the new string formed after each operation is also a valid word in the language. This %%game of words is described in the novel ``Alice in Wonderland"~\cite{alice}. It is easy to see that the value of $\langle %%l \rangle$ at $\theta = 1$ specifies the number of edit operations that needs to be performed on an average. On the other %%hand $D$ specifies the worst case bound on the number of edit operations for this game. Note that $D$ is much greater %%than $\langle l \rangle$. Moreover, though the SpellNets of Hindi and Bengali are denser than that of Enlish, the values of %% $\langle l \rangle$ and $D$ are comparable for the networks.

%%%%%%%%%%%%%%%%%%%%%%%%%%%%%%%%%%%%%%%%%%%%%%%%%%%%%%%%%%%%%%%%%%%%%%%%%%%%%%%%%%%%%%%%%%%%%%%%%%%%%%%%%
%%%%%%%%%%%%%%%%%%%%%%%%%%%%%%%%%%%% SECTION %%%%%%%%%%%%%%%%%%%%%%%%%%%%%%%%%%%%%%%%%%%%%%%%%%%%%%%%%%%%
%%%%%%%%%%%%%%%%%%%%%%%%%%%%%%%%%%%%%%%%%%%%%%%%%%%%%%%%%%%%%%%%%%%%%%%%%%%%%%%%%%%%%%%%%%%%%%%%%%%%%%%%%
%%%%%%%%%%%%%%%%%%%%%%%%%%%%%%%%%%%%%%%%%%%%%%%%%%%%%%%%%%%%%%%%%%%%%%%%%%%%%%%%%%%%%%%%%%%%%%%%%%%%%%%%%

\section{Conclusion}\label{sec:con}

In this work, we have proposed the network of orthographic neighbors of words or the SpellNet and studied the structure of the same across three languages. We have also made an attempt to relate some of the topological properties of SpellNet to spelling error distribution and hardness of spell-checking in a language. The important observations of this study are summarized below. 
\begin{itemize}
\item The probability of RWE in a language can be equated to the average weighted degree of SpellNet. This probablity is highest in Hindi followed by Bengali and English.
\item In all the languages, the words that are more prone to undergo an RWE are more likely to be misspelt. Effectively, this makes RWE correction very hard. 
\item The hardness of NWE correction correlates with the weighted clustering coefficient of the network. This is highest for Hindi, followed by Bengali and English.
\item The basic topology of SpellNet seems to be an invariant across languages. For example, all the networks feature exponential degree distribution, high clustering, assortative mixing with respect to degree and node weight, small world effect and positive correlation between degree and node weight, and CC and degree. However, the networks vary to a large extent in terms of the actual values of some of these metrics.
\end{itemize}

Arguably, the language-invariant properties of SpellNet can be attributed to the organization of the human mental lexicon (see~\cite{kapatsinski} and references therein), self-organization of orthographic systems and certain properties of edit distance measure. The differences across the languages, perhaps, are an outcome of the specific orthographic features, such as the size of the alphabet. Another interesting observation is that the phonemic nature of the orthography strongly correlates with the difficulty of spell-checking. Among the three languages, Hindi has the most phonemic and English the least phonemic orthography. This correlation calls for further investigation.

Throughout the present discussion, we have focussed on spell-checkers that ignore the context; consequently, many of the aforementioned results, especially those involving spelling correction,  are valid only for context-insensitive spell-checkers. Nevertheless, many of the practically useful spell-checkers incorporate context information and the current analysis on SpellNet can be extended for such spell-checkers by conceptualizing a network of words that capture the word co-occurrence patterns~\cite{chris}. The word co-occurrence network can be superimposed on SpellNet and the properties of the resulting structure can be appropriately analyzed to obtain similar bounds on hardness of context-sensitive spell-checkers. We deem this to be a part of our future work. Another way to improve the study could be to incorporate a more realistic measure for the orthographic similarity between the words. Nevertheless, such a modification will have no effect on the analysis technique, though the results of the analysis may be different from the ones reported here.

%%%%%%%%%%%%%%%%%%%%%%%%%%%%%%%%%%%%%%%%%%%%%%%%%%%%%%%%%%%%%%%%%%%%%%%%%%%%%%%%%%%%%%%%%%%%%%%%%%%%%%%%%
%%%%%%%%%%%%%%%%%%%%%%%%%%%%%%%%%%%% SECTION %%%%%%%%%%%%%%%%%%%%%%%%%%%%%%%%%%%%%%%%%%%%%%%%%%%%%%%%%%%%
%%%%%%%%%%%%%%%%%%%%%%%%%%%%%%%%%%%%%%%%%%%%%%%%%%%%%%%%%%%%%%%%%%%%%%%%%%%%%%%%%%%%%%%%%%%%%%%%%%%%%%%%%
%%%%%%%%%%%%%%%%%%%%%%%%%%%%%%%%%%%%%%%%%%%%%%%%%%%%%%%%%%%%%%%%%%%%%%%%%%%%%%%%%%%%%%%%%%%%%%%%%%%%%%%%%
\section*{Appendix A: Derivation of the Probability of RWE}

We take a {\em noisy channel approach}, which is a common technique in NLP (for example~\cite{ibm}), including spellchecking~\cite{spellchk_nc}. Depending on the situation. the channel may model typing or OCR errors. Suppose that a word $w$, while passing through the channel, gets transformed to a word $w'$. Therefore, the aim of spelling correction is to find the $w^* \in \Lambda$ (the lexicon), which maximizes $p(w^*|w')$, that is
\begin{equation}\label{nc} 
\begin{array}{c}\\argmax\\^{w\in \Lambda}\end{array}p(w|w')= \begin{array}{c}\\argmax\\^{w\in \Lambda}\end{array}p(w'|w)p(w)
\end{equation}
The likelihood $p(w'|w)$ models the noisy channel, whereas the term $p(w)$ is traditionally referred to as the
{\em language model} (see~\cite{jurafsky} for an introduction). In this equation, as well as throughout this discussion, we shall assume a unigram language model, where $p(w)$ is the normalized frequency of occurrence of $w$ in a standard corpus. 

We define the probability of RWE for a word $w$, $p_{rwe}(w)$, as follows
\begin{equation}\label{rwe1}
p_{rwe}(w) = \sum_{\begin{array}{c}^{w' \in \Lambda}\\ ^{w \ne w'}\end{array}} p(w'|w)
\end{equation}
Stated differently, $p_{rwe}(w)$ is a measure of the probability that while passing through the channel, $w$ gets transformed into a form $w'$, such that $w'\in\Lambda$ and $w'\ne w$. The probability of RWE in the language, denoted by $p_{rwe}(\Lambda)$, can then be defined in terms of the probability $p_{rwe}(w)$ as follows.
\begin{eqnarray}\label{rwe2}
p_{rwe}(\Lambda) = \sum_{w \in \Lambda}p_{rwe}(w)p(w) \\
 = \sum_{w \in \Lambda}\sum_{\begin{array}{c}^{w' \in \Lambda}\\ ^{w \ne w'}\end{array}} p(w'|w)p(w) \label{rwe}\nonumber 
\end{eqnarray}

In order to obtain an estimate of the likelihood $p(w'|w)$, we use the concept of {\em edit distance} (also known as Levenstein distance~\cite{levenstein}). We shall denote the edit distance between two words $w$ and $w'$ by $ed(w,w')$. If we assume that the probability of a single error (i.e. a character deletion, substitution or insertion) is $\rho$ and errors are independent of each other, then we can approximate the likelihood estimate as follows.
\begin{equation}\label{exp}
p(w'|w) = \rho^{ed(w,w')}
\end{equation}
Exponentiation of edit distance is a common measure of word similarity or likelihood (see for example~\cite{bailey:01}).

Substituting for $p(w'|w)$ in Eqn~\ref{rwe2}, we get
\begin{equation}
p_{rwe}(\Lambda) = \sum_{w \in \Lambda}\sum_{\begin{array}{c}^{w' \in \Lambda}\\ ^{w \ne w'}\end{array}}\rho^{ed(w,w')}p(w)
\end{equation}

%%%%%%%%%%%%%%%%%%%%%%%%%%%%%%%%%%%%%%%%%%%%%%%%%%%%%%%%%%%%%%%%%%%%%%%%%%%%%%%%%%%%%%%%%%%%%%%%%%%%%%%%%
%%%%%%%%%%%%%%%%%%%%%%%%%%%%%%%%%%%% SECTION %%%%%%%%%%%%%%%%%%%%%%%%%%%%%%%%%%%%%%%%%%%%%%%%%%%%%%%%%%%%
%%%%%%%%%%%%%%%%%%%%%%%%%%%%%%%%%%%%%%%%%%%%%%%%%%%%%%%%%%%%%%%%%%%%%%%%%%%%%%%%%%%%%%%%%%%%%%%%%%%%%%%%%
%%%%%%%%%%%%%%%%%%%%%%%%%%%%%%%%%%%%%%%%%%%%%%%%%%%%%%%%%%%%%%%%%%%%%%%%%%%%%%%%%%%%%%%%%%%%%%%%%%%%%%%%%

\bibliographystyle{acl}
\bibliography{tg}

\end{document}